\documentclass[twocolumn,showpacs,preprintnumbers,amsmath,amssymb,superscriptaddress]{revtex4}
\usepackage[]{umlaut,latexsym}
\usepackage{amssymb}
\usepackage[USenglish]{babel}
\usepackage{graphicx}
\usepackage{dcolumn}
\usepackage[dvips]{color}
\usepackage{ulem}

\usepackage{float}
\floatstyle{boxed} 

\begin{document}

\title{\bf Anisotropic ferromagnetism in carbon doped zinc oxide from first-principles studies}

\author{Sanjeev~K.~Nayak} \email{sanjeev.nayak@uni-due.de}
\affiliation{Faculty of Physics and Center for Nanointegration (CENIDE), University of Duisburg-Essen, 47048 Duisburg, Germany}
\author{Markus~E.~Gruner}
\affiliation{Faculty of Physics and Center for Nanointegration (CENIDE), University of Duisburg-Essen, 47048 Duisburg, Germany}
\author{Sung~Sakong}
\affiliation{Faculty of Physics and Center for Nanointegration (CENIDE), University of Duisburg-Essen, 47048 Duisburg, Germany}
\author{Shreekantha~Sil}
\affiliation{Department of Physics, Visva Bharati University, Santiniketan 731235, West Bengal, India}
\author{Peter~Kratzer}
\affiliation{Faculty of Physics and Center for Nanointegration (CENIDE), University of Duisburg-Essen, 47048 Duisburg, Germany}
\author{Surjyo~N.~Behera\footnotemark\footnotetext{Prof.~Behera passed away on 14.12.2011 during the preparation of the manuscript.}}
\affiliation{School of Electrical Sciences, Indian Institute of Technology, Bhubaneswar 751013, India}
\altaffiliation{Emeritus Professor at National Institute of Science and Technology, Palur Hills, Berhampur 761008, India}
\author{Peter~Entel}
\affiliation{Faculty of Physics and Center for Nanointegration (CENIDE), University of Duisburg-Essen, 47048 Duisburg, Germany}

\date{\today}

\begin{abstract}
A density functional theory study of substitutional carbon impurities in ZnO has been performed, 
using both the generalized gradient approximation (GGA) and a hybrid functional (HSE06) as 
exchange-correlation functional. It is found that the non-spinpolarized C$_\mathrm{Zn}$ impurity 
is under almost all conditions thermodynamically more stable than the C$_\mathrm{O}$ impurity 
which has a magnetic moment of $2\mu_{\mathrm{B}}$, with the exception of very O-poor and C-rich 
conditions. This explains the experimental difficulties in sample preparation in order to realize 
$d^{0}$-ferromagnetism in C-doped ZnO. From GGA calculations with large 96-atom supercells, 
we conclude that two C$_\mathrm{O}$-C$_\mathrm{O}$ impurities in ZnO interact ferromagnetically, 
but the interaction is found to be short-ranged and anisotropic, much stronger within the hexagonal 
$ab$-plane of wurtzite ZnO than along the $c$-axis. This layered ferromagnetism is attributed to 
the anisotropy of the dispersion of carbon impurity bands near the Fermi level for C$_{\mathrm{O}}$ 
impurities in ZnO. From the calculated results, we derive that a C$_{\mathrm{O}}$ concentration 
between 2\% and 6\% should be optimal to achieve $d^{0}$-ferromagnetism in C-doped ZnO. 
\end{abstract}
\maketitle

\section{Introduction}

Ferromagnetism is a quantum phenomenon which arises from the long range ordering of interacting 
magnetic moments in a solid. The richness of magnetic properties in solids is due to hybridization 
of $d$- and $f$-orbitals which is decisive for magnetic moment formation and the way individual 
moments interact with each other in the solid. This is one of the reasons why magnetic 
semiconductors or diluted magnetic semiconductors (DMS) are traditionally conceived as 
semiconducting materials with transition metal ions as dopant. Numerous studies of DMS, 
both experimental and theoretical 
(see, e.g. Refs.~\cite{DMSreview01,DMSreview02,DMSreview03,Dietl-2010, Ney-2010}) 
have come up with varying conclusions. In ZnO, substituting the cation by an isovalent 
transition metal, e.g. cobalt, introduces magnetic moments, but in itself does not lead to 
ferromagnetic coupling between these moments~\cite{Ney-2010}. Hence, complex strategies, 
for instance co-doping, and elaborate growth techniques seem to be necessary to fabricate 
a DMS based on ZnO as host material. 

Recently there are reports on magnetism in semiconducting materials without any 
$d$- or $f$-elements~\cite{Kenmochi-2004,Venkatesan-2004,Ivanovskii-2007,Volnianska-2010}. 
This is commonly known as $d^{0}$-magnetism and discussed for different classes of materials, 
starting from pure carbon based materials like graphite, graphene, and non-metallic 
nanoparticles~\cite{Esquinazi-2003,Makarova-2006}. The common feature in this class 
of materials is that magnetism involves the 2$p$ orbitals. The origin of magnetic moment 
is attributed to incomplete cancellation of majority and minority spin electronic 
contribution due to the Hund's type exchange. Magnetic moments have been reported for 
almost all ionic semiconductors in the presence of intrinsic defects, such as cation 
vacancies, or in the presence of impurities related to anion substitution. In the following 
we discuss reports of magnetism in semiconductors with impurities at the anion site.

The formation of a localized magnetic moment, which is a pre-requisite for $d^{0}$-magnetism, 
depends on the relative strength of electronegativity between the dopant element and the 
anion of the host semiconductor. If the bond between the dopant and the cation is weaker 
than the native bond of the semiconductor, this leads to localized atomic-like 2$p$ orbitals 
of the dopant and a stable spin-polarized state. On the other hand, if the strength of the 
bond is stronger than the native bond of the semiconductor, this will cause delocalization 
of the dopant $2p$ orbitals due to strong hybridization with the cations. Consequently, there 
is reduced or vanishing spin-polarization in the system~\cite{Yang-2010-1}. Hence, substituting 
the anion site by an element with smaller electronegativity can introduce spin-polarized 
states in the gap which can then mediate the magnetic interaction through the double-exchange 
mechanism as discussed by Mavropoulos {\it et al.}~\cite{Mavropoulos-2009}. 

By means of DFT calculations, Peng {\it et al.}~\cite{Peng-2009-3} have shown that many 
elements in the second row of the periodic table (X = B, N, C) when substituted at the 
anion site can develop a spin polarized solution in AlN (X$_{\mathrm{N}}$) and ZnO 
(X$_{\mathrm{O}}$). However, the nature of the magnetic correlation in case of 
B$_{\mathrm{O}}$ and N$_{\mathrm{O}}$ is still under 
debate~\cite{Peng-2009-3,Elfimov-2007,Shen-2008,Peng-2009-1}. Specifically for ZnO, 
Adeagbo {\it et al.} found that both N$_{\mathrm{O}}$ and N$_{\mathrm{Zn}}$ develop 
spin-polarization with almost 1~$\mu_{\mathrm{B}}$/N in ZnO~\cite{Adeagbo-2010}. 
Lyons~{\it et~al.} have studied N$_{\mathrm{O}}$ in ZnO by hybrid-functional calculations 
and observed that N$_{\mathrm{O}}$ defects create a deep impurity state~\cite{Lyons-2009}. 
These  states, if partially filled, can be spin-polarized. Also for C$_{\mathrm{O}}$, 
DFT calculations predicted a magnetic moment for the neutral or singly charged 
impurity~\cite{Sakong-2011}. There are experimental reports of room temperature 
ferromagnetism in carbon-doped ZnO films grown by pulsed-laser deposition~\cite{Pan-2007,Zhou-2008}. 
Measurements of the anomalous Hall effect have shown that the ferromagnetism is 
mediated by the charge carriers in ZnO. Remarkably, results have been reported both 
for samples showing n-type conductivity~\cite{Pan-2007} or p-type conductivity~\cite{Herng-2009}.
Moreover, Yi~{\it et~al.} reported on room-temperature ferromagnetism in carbon 
doped ZnO films when co-doped with nitrogen~\cite{Yi-2009}. 

In this work, we select ZnO as a host material for the study of $d^{0}$ magnetism. 
Impurities and defects in ZnO have been an active topic of research~\cite{Mccluskey-2009,Ogale-2010}. 
Here we focus on the magnetic properties of ZnO with carbon as substitutional impurity 
(on the zinc site denoted by C$_\mathrm{Zn}$, and on the oxygen site denoted by C$_\mathrm{O}$) 
studied with the help of density functional theory (DFT) calculations. The paper is arranged 
as follows: In the next Section we present technical details of our DFT calculations. 
The Results and Discussion Section discusses the stability of carbon impurities in ZnO 
by calculating their formation energies. The distance-dependent exchange interactions of 
C$_\mathrm{O}$ in ZnO are obtained from the total energy differences of ferromagnetic 
and antiferromagnetic states. Finally, we conclude on the possible connection 
between C impurities and the ferromagnetism in ZnO found in experiments. 

\section{\label{computationalmethods} Computational Details}

Electronic structure calculations are performed using the plane-wave pseudopotential 
DFT method as implemented in the Vienna {\it ab-initio} simulation package 
(VASP)~\cite{VASP01,VASP02}. For studying a substitutional impurity in ZnO, 
the system is modeled as a supercell consisting of periodic repetitions of the 
primitive wurtzite cell. Since we use different sizes of supercells, we stick to 
a particular nomenclature in our discussions. We refer to the size of the supercell 
with periodic extention $n_{x}\negmedspace\times\negmedspace 
n_{y}\negmedspace\times\negmedspace n_{z}$ as ``S$n_{x}n_{y}n_{z}$". 
The electronic structure of homogeneously distributed impurites has been studied with 
the relatively small supercell S222, see Fig. 1(c), while to study impurity pairs 
in ZnO we employ two kinds of large supercells, namely S622 and S226 
(Fig. 1(d) and (e), respectively. This is necessary as the wurtzite crystal 
structure is non-centrosymmetric and the bond lengths are different along the 
hexagonal plane and along the $c$-direction. Essentially, S622 intends to scan 
the interactions along the hexagonal plane of the lattice and the S226 scans the 
interaction along the $c$-axis of the lattice. 

\begin{figure}[t]
\centering{
\includegraphics[scale=0.07]{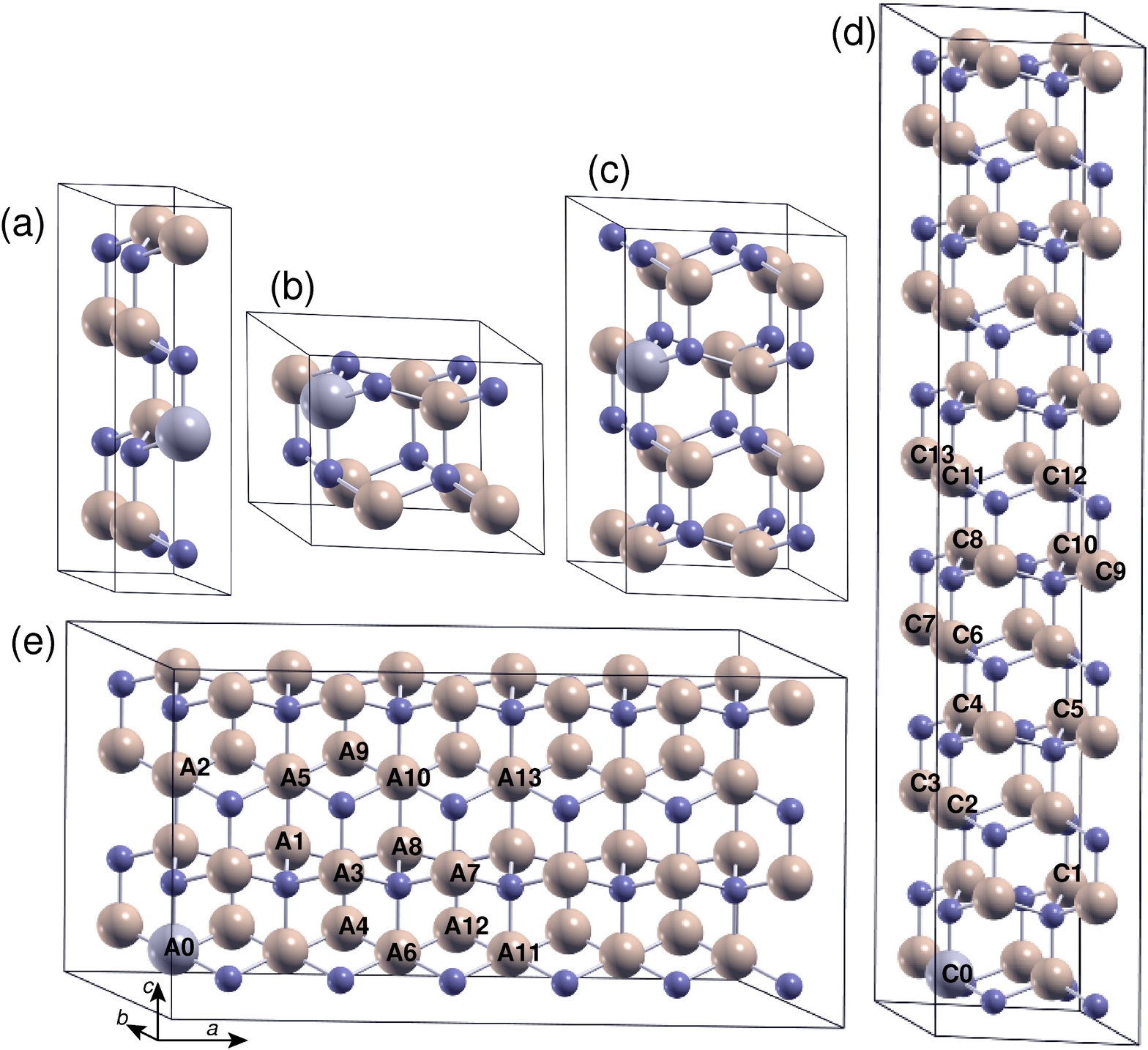}
}
\caption{\label{four_supercells} (Color online) Supercells and impurity 
configurations used in the present study. The small (blue) spheres represent Zn, 
and among the large spheres, the darker ones represent carbon on the O sites 
while light spheres represent oxygen atoms in the ZnO lattice. The supercells 
S122 (a), S221 (b) and S222 (c) contain a single impurity and are used to 
model impurity concentrations of 12.5\% for S122 and S221 and 6.25\% for S222. 
The C$_{\mathrm{O}}$-C$_{\mathrm{O}}$ interaction is studied with two 
C$_{\mathrm{O}}$ in S226 (d) and S622 (e). One of them fixed at the 
position C0 and A0 in S226 and S622, respectively, and another placed on one 
of the C1--C13/A1--A13 sites, which corresponds to a concentration of 4.17\%.}
\end{figure}

The calculations have been performed using a plane-wave energy cutoff of 400 eV. 
We have used the generalized gradient approximation~(GGA) with the PW91 
parameterization~\cite{PW91} for the exchange-correlation potential and the 
ion-electron interactions are treated with the projector-augmented wave~(PAW) 
method~\cite{VASP03}. The lattice constants of the primitive lattice of ZnO 
is taken to be $a$ = 3.29 {\AA}, $c/a$ = 1.606 and the relative shift of 
Zn and O planes, $u$~=~0.380, which were found to be the minimum energy 
lattice parameters in the GGA calculations~\cite{Nayak-2009}. The doped 
supercell configuration is relaxed to avoid any strain until the structural 
energy is converged to $10^{-6}$ eV. At this level of tolerance, the forces 
in the system are found to be below 
$3\negmedspace\times\negmedspace 10^{-4}~\mathrm{eV~\AA^{-1}}$. 
The relaxation is performed with a $\Gamma$-centered Monkhorst-Pack $k$-points 
grid of $3 \negmedspace\times\negmedspace 5 \negmedspace\times\negmedspace 5$ 
and $5\negmedspace\times\negmedspace 5\negmedspace\times\negmedspace 3$ for 
supercells S622 and S226, respectively. 

Part of these results have been cross-checked with the full-potential local 
orbital (FPLO) minimum-basis code, which uses localized basis sets~\cite{Koepernik-1999}.  
DFT within GGA gives reliable structural parameters for ZnO but has limitation 
in predicting the right optical band gap, which is with a value of 0.69 eV grossly
underestimated in comparison to the experimental value of 3.4 eV~\cite{Nayak-2009}. 
In addition, the localization of Zn $d$-bands is not properly described by GGA because 
the hybridization of Zn $d$ and O $p$ bands is overestimated. Thus, the Zn $d$ bands 
are located almost 3 eV too close to the Fermi level as compared to experimental 
observation~\cite{Powell-1971}. One way to overcome the problem is through the 
introduction of hybrid functionals where the electronic exchange potential is 
mixed with some percentage of Hartree-Fock exchange, while the electronic correlation 
potential is entirely  taken from the GGA. We have employed the screened 
hybrid-functionals to ascertain the electronic structure of a single impurity. 
The hybrid-functionals are of the type suggested by Heyd, Scuseria and 
Ernzerhof~\cite{HSE-2003}, with the the exact (Hartree-Fock) exchange and the 
GGA exchange in the ratio 1:3~\cite{HSE-alpha}. We have used the HSE06 functional 
for our studies~\cite{HSE-03-06}.

\section{Results and discussion}

The stability of an impurity configuration can be judged from its formation energy. 
Since substitution changes the number of atomic species, the formation energy 
must be calculated with  reference to thermodynamic reservoirs characterized by 
their chemical potentials. In practice, the values of these chemical potentials 
reflect the experimental conditions during sample preparation. The formation energy, 
E$_{f}$, is thus defined as 
\begin{equation} \label{eq.formationenergy}
E_{f} = E_{\mathrm{Tot}} - E_{\mathrm{ZnO}} - \sum_{i}n_{i}\mu_{i},
\end{equation}
where 
$E_{\mathrm{Tot}}$ is the total energy of the supercell with impurity, 
$E_{\mathrm{ZnO}}$ is the energy of pure ZnO host with the equivalent number of 
stoichiometric ZnO units, $\mu_{i}$ is the chemical potential of a species and 
$n_{i}$ is the change of number of the corresponding species in the supercell.
In thermodynamic equilibrium with bulk ZnO, the condition 
\begin{equation}\label{znogrowth}
\mu_{\mathrm{ZnO}} = \mu_{\mathrm{O}}+\mu_{\mathrm{Zn}}. 
\end{equation}
must always be satisfied. This leaves us with just one unknown, which we choose 
to be $\mu_{\mathrm{O}}$. However, an interval for the values of $\mu_{\mathrm{O}}$ 
corresponding to physically meaningful growth conditions can be provided:
The upper bound of the chemical potential, corresponding to O-rich growth conditions, 
is taken from the oxygen molecule, $\mu_{\mathrm{O}} = \frac{1}{2}E_{\mathrm{O_{2}}}$. 
Oxygen-poor conditions (or, equivalently, Zn-rich conditions) correspond to the 
equilibrium with metallic bulk Zn, $\mu_{\mathrm{Zn}} = E_{\mathrm{Zn}}^{\mathrm{bulk}}$, 
and hence, from Eq.~(\ref{znogrowth}), 
$\mu_{\mathrm{O}} = \mu_{\mathrm{ZnO}} - E_{\mathrm{Zn}}^{\mathrm{bulk}}$. 

\begin{table*}[]
\caption{\label{tab:formationenergy2}
Formation energy of C$_{\mathrm{O}}$ and C$_{\mathrm{Zn}}$ calculated from 
Eq.~(\ref{eq.formationenergy}). The data show that the C$_{\mathrm{Zn}}$ impurity is 
thermodynamically more stable than the C$_{\mathrm{O}}$ impurity.}
\begin{tabular}{llllll}  
\hline
\hline
        & & & carbon rich & & carbon poor \\
\hline
O-poor, &C$_\mathrm{O}$   & & -4.558                                & & 4.070 \\
        &C$_\mathrm{Zn}$  & & -4.576                                & & 4.052 (4.0 Ref.~\cite{Tan-2007}) \\
O-rich, &C$_\mathrm{O}$   & & -1.569 (0.4 Ref.~\cite{Sakong-2011})  & & 7.059 (9.4 Ref.~\cite{Sakong-2011}) \\
        &C$_\mathrm{Zn}$  & & -7.564 (-6.2 Ref.~\cite{Sakong-2011}) & & 1.064 (2.8 Ref.~\cite{Sakong-2011}, 2.2 Ref.~\cite{Tan-2007})\\
\hline
\hline
\end{tabular}
\end{table*}

\subsection{C$_\mathrm{Zn}$ and C$_\mathrm{O}$ impurities}

\begin{figure}[t]
\centering{
\includegraphics[scale=0.38]{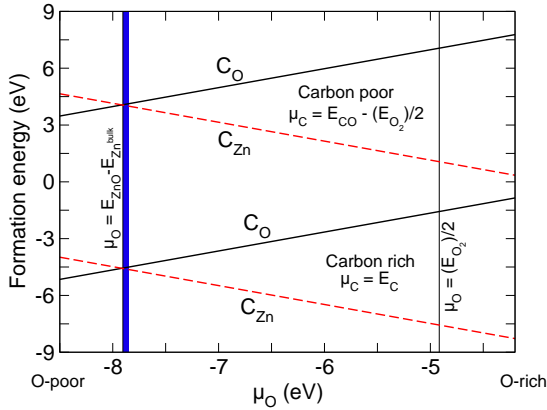}}
\caption{\label{phaseplot} (Color online) The formation energy of C$_{\mathrm{O}}$
and C$_{\mathrm{Zn}}$ defects as a function of $\mu_{\mathrm{O}}$. The thick vertical
line (blue colored line) shows the narrow range where the C$_{\mathrm{O}}$ impurity 
can compete with C$_{\mathrm{Zn}}$.}
\end{figure}

The formation energies of carbon at the Zn-site (C$_{\mathrm{Zn}}$) and at the 
O-site (C$_{\mathrm{O}}$) obtained from calculations with a 
$3 \negmedspace\times\negmedspace 3\negmedspace\times\negmedspace 3$
supercell of ZnO are summarized in Table~\ref{tab:formationenergy2} for two 
different environments. The limiting chemical potentials of the C-rich and 
for C-poor cases are taken as $\mu_{\mathrm{C}} = E_{\mathrm{C}}$ and 
$\mu_{\mathrm{C}} = E_{\mathrm{CO}} - \frac{1}{2}E_{\mathrm{O}_{2}}$, respectively, 
as described in Ref.~\cite{Sakong-2011}. We note that the deviation of the values 
as compared to Ref.~\cite{Sakong-2011} is due to the different exchange-correlation 
functionals used in both calculations. The values obtained by 
Tan~{\it et al.}~\cite{Tan-2007} should not be directly compared to ours because 
they have used the total energies of single atoms  as reference chemical potentials. 
In Fig.~\ref{phaseplot}, the formation energies which we obtain for C$_{\mathrm{O}}$
and C$_{\mathrm{Zn}}$ impurities are plotted as a function of oxygen chemical potential. 
Our formation energies suggest that C$_\mathrm{O}$ and C$_\mathrm{Zn}$ impurities 
are energetically favorable only under carbon-rich conditions (negative values in 
the formation energy). Also, the formation energy of C$_\mathrm{Zn}$ is smaller 
than the formation energy of C$_\mathrm{O}$ in all chemical environments.

We find that the formation of C$_{\mathrm{O}}$ is energetically favorable only 
in a narrow range of $\mu_{\mathrm{O}}$ shown as the thick (blue) vertical line. 
This tiny region corresponds to an equilibrium with metallic zinc, i.e., extremely 
O-poor conditions. Hence stabilizing the C$_{\mathrm{O}}$ defect in ZnO is difficult 
under common experimental growth conditions of high oxygen pressure where 
$\mu_{\mathrm{O}}$ approaches the value corresponding to molecular oxygen.

\begin{figure}[t]
\centering{
\includegraphics[scale=0.20]{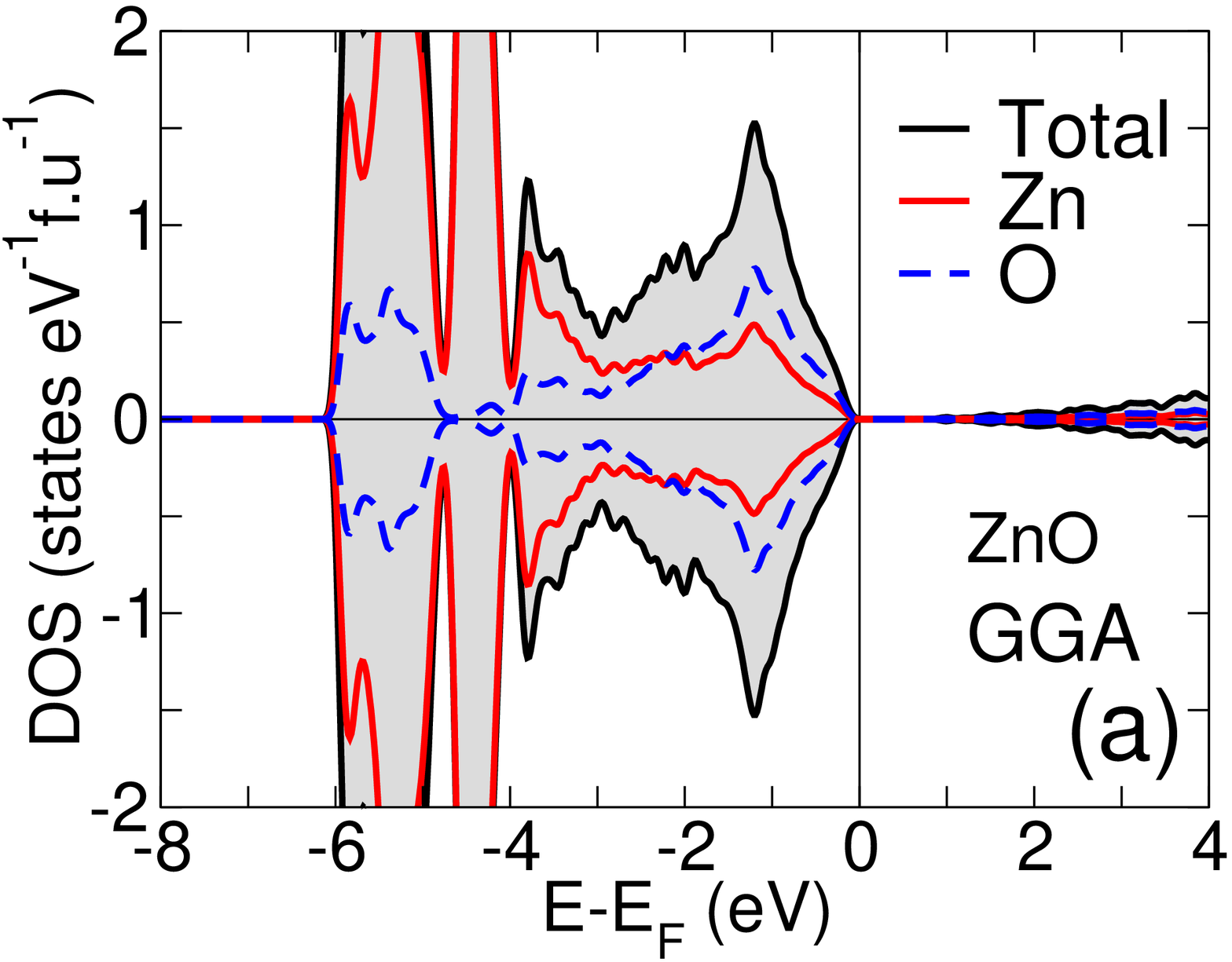}
\includegraphics[scale=0.20]{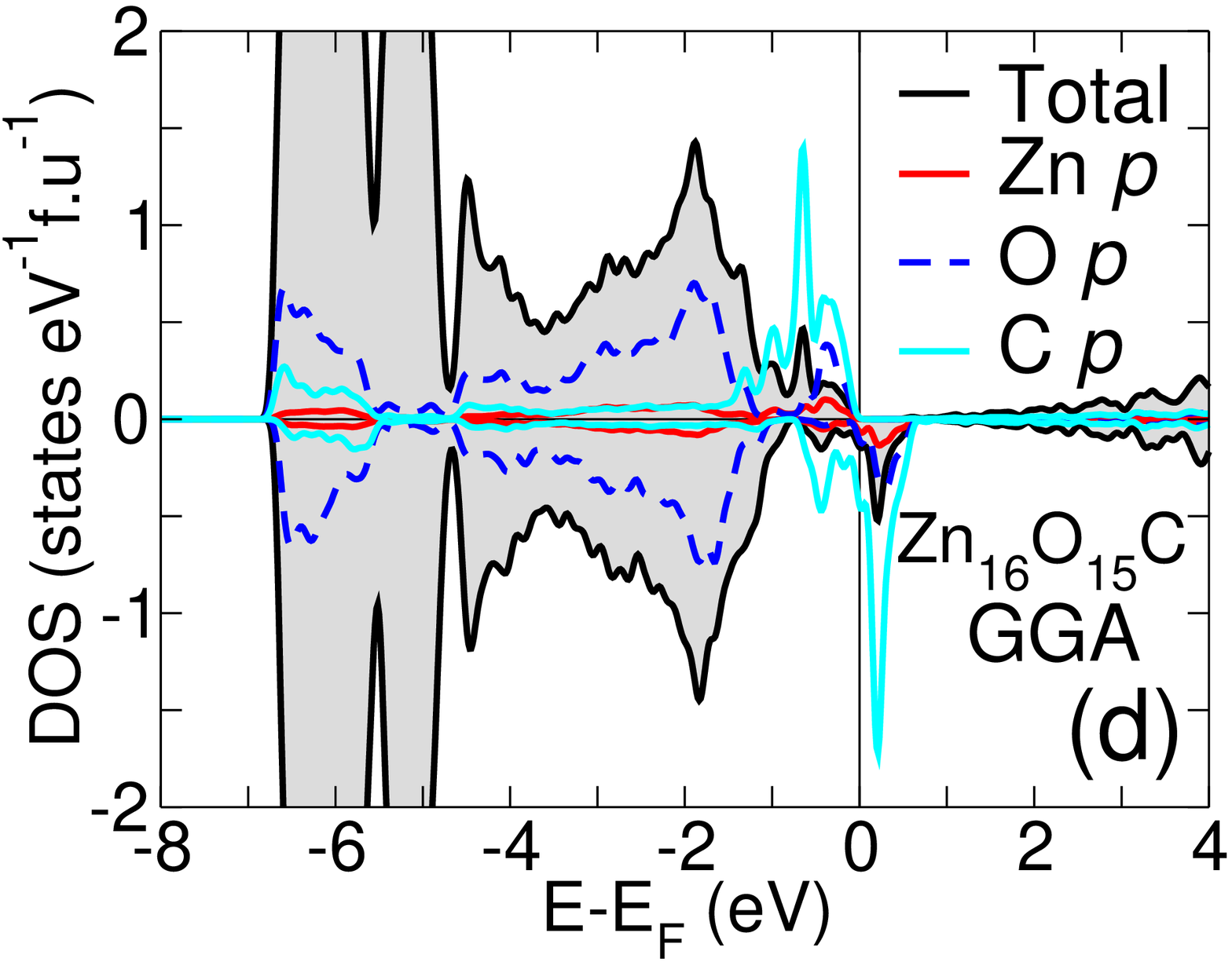}
\includegraphics[scale=0.20]{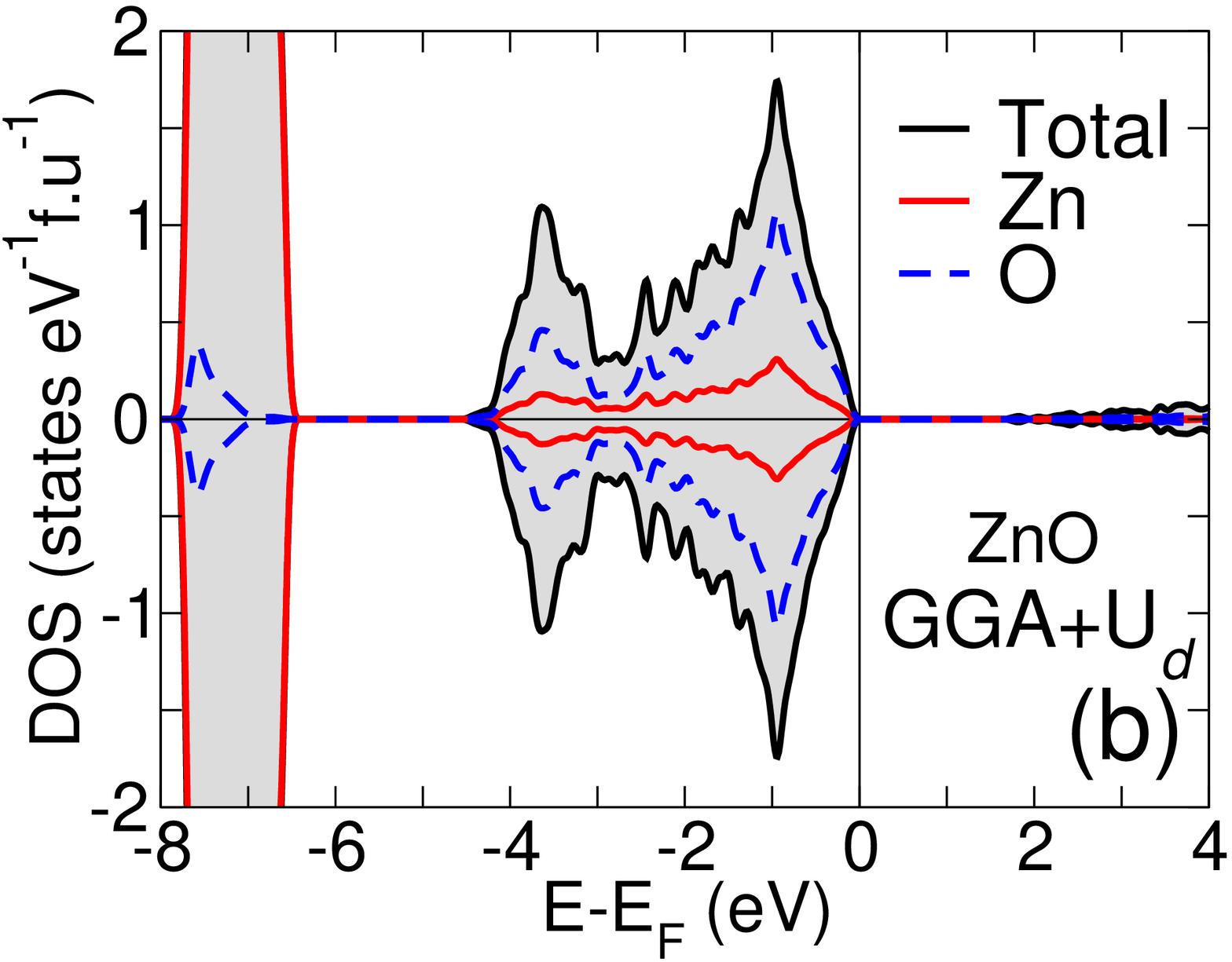}
\includegraphics[scale=0.20]{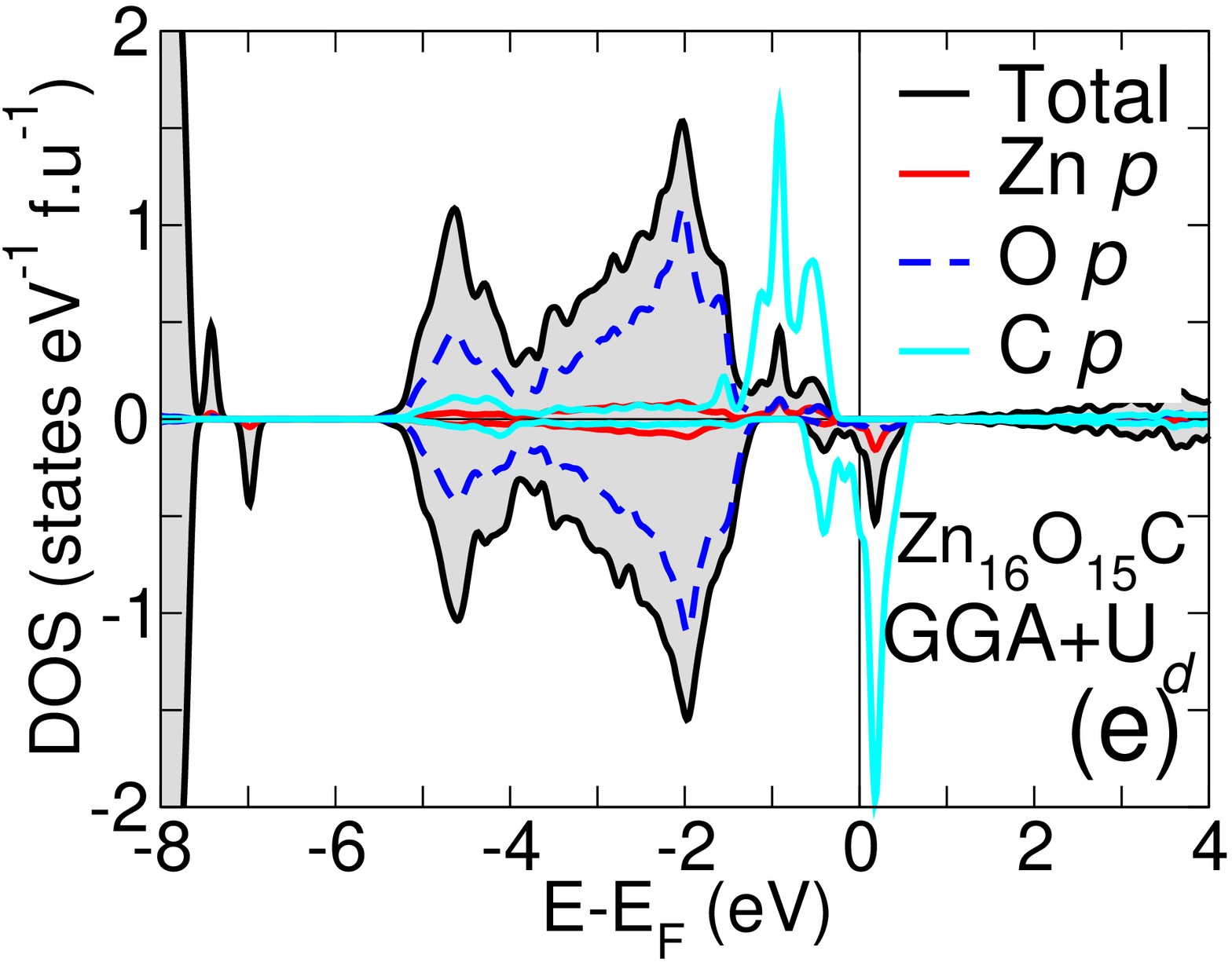}
\includegraphics[scale=0.20]{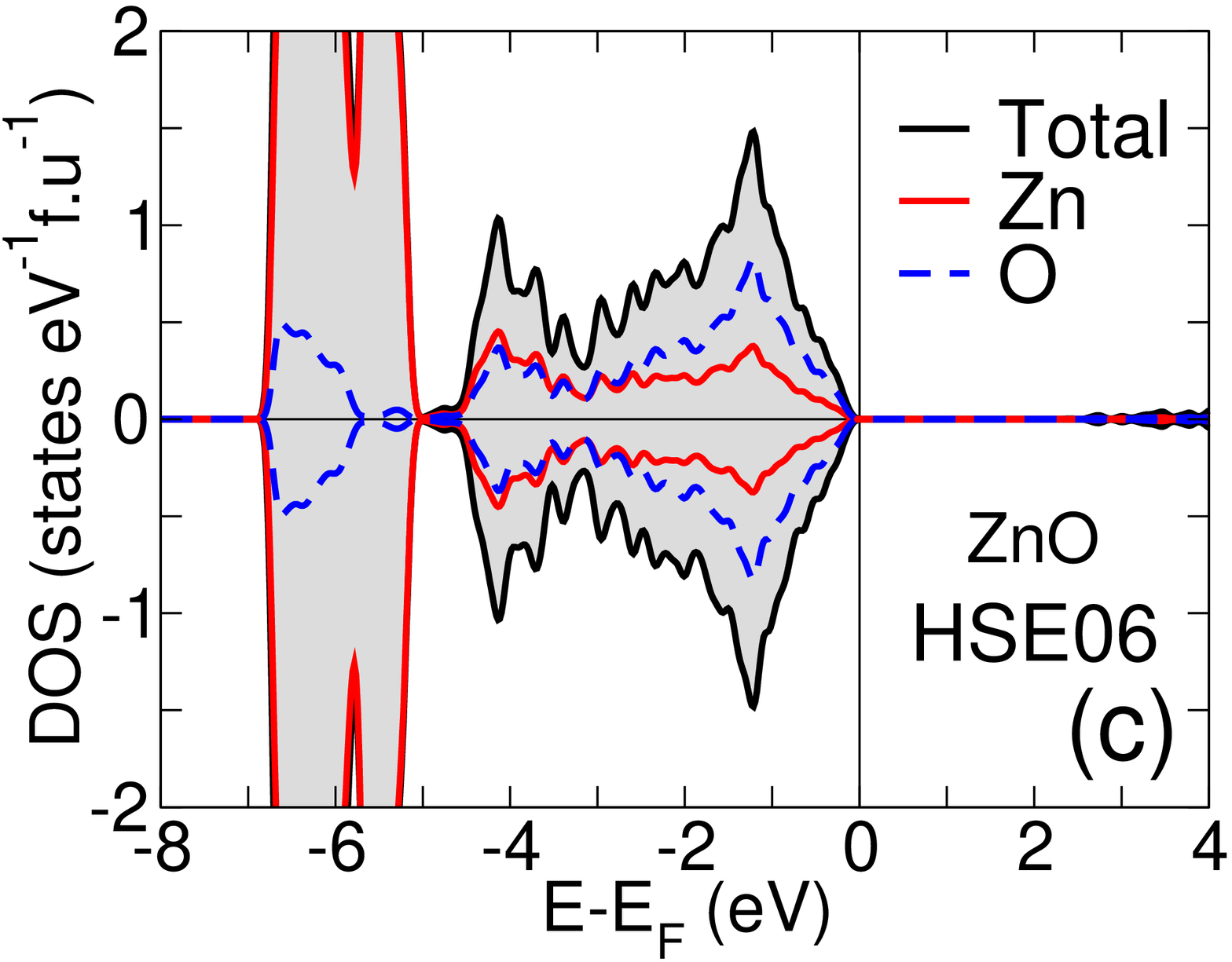}
\includegraphics[scale=0.20]{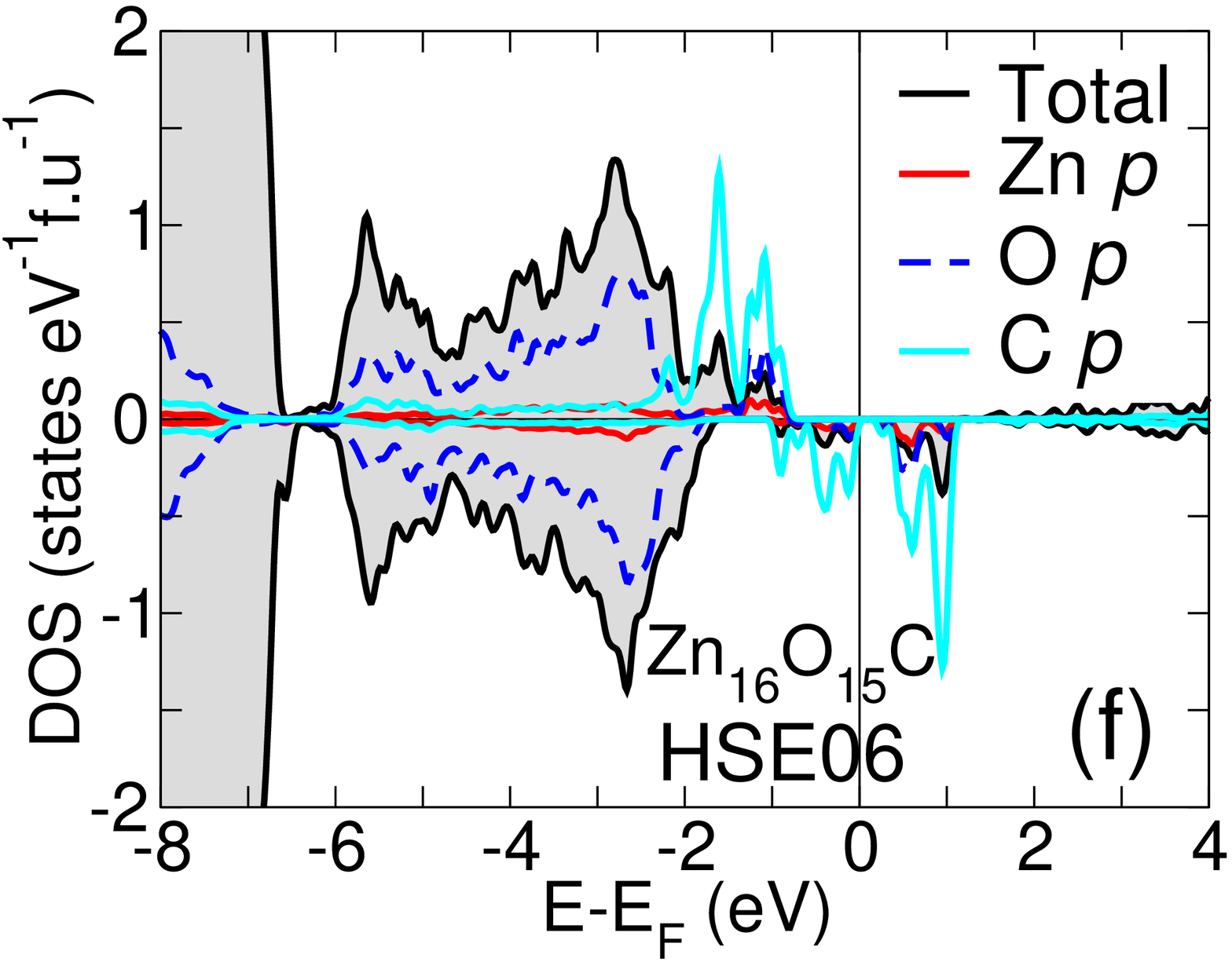}
}
\caption{\label{fig:dosplot} (Color online) Density of states (DOS) of pure 
ZnO ((a), (b), (c)) and C$_{\mathrm{O}}$ impurity in ZnO ((d), (e), (f)).
The DOS in different rows are obtained from GGA, GGA+$U_{d}$ on Zn and from
HSE06 functional. The spin-up DOS and the spin-down DOS are plotted as positive 
and negative $y$-axis. The DOS of C$_{\mathrm{O}}$ impurity in ZnO in S222 geometry 
consists of the atomic contribution of nearest neighbor Zn, O to C in units of 
states/atom, while the total DOS is scaled to the number of formula units of ZnO.}
\end{figure}

\begin{figure}[]
\centering{\includegraphics[scale=0.055]{Fig_04.eps}}
\caption{\label{fig:chgplot} (Color online) Effective spin density 
($\Delta\rho = \rho_{\uparrow}-\rho_{\downarrow}$) plotted along the plane perpendicular 
to $c$-axis and passing through C$_{\mathrm{O}}$. The light (magenta), dark (turquoise) 
and black colored balls are used to represent Zn, O and C atoms~\cite{Xcrysden}. 
The isolines (and also the color gradient) are plotted from 0.003 $e\mathrm{\AA}^{-3}$ 
to 0.028 $e\mathrm{\AA}^{-3}$ at intervals of 0.005 $e\mathrm{\AA}^{-3}$. One can clearly 
observe that the spin-polarization arising from C$_{\mathrm{O}}$ extends to relatively 
large spatial region.}
\end{figure}

A single C$_{\mathrm{O}}$ in a ZnO supercell leads to a spin-polarized solution with 
an integral magnetic moment of 2$\mu_{B}$/C as discussed in the literature 
previously~\cite{Lin-2010,Pan-2007,Yang-2010-1}. The spin polarized solution is lower 
in energy than the non-spin polarized one by a difference of 0.215 eV. The corresponding 
DOS of C$_{\mathrm{O}}$ as single impurity in supercell S222 is shown in Fig.~\ref{fig:dosplot}
(d), (e) and (f), employing the GGA, GGA+$U_{d}$ and HSE06 functionals. For comparison, 
the DOS of pure ZnO obtained from each method is shown in 
Fig.~\ref{fig:dosplot}~(a), (b) and (c). In GGA+$U_{d}$, the value $U_{d}=7.5$~ eV is 
applied to the (fully occupied) Zn $d$-orbitals. We find a magnetic solution for 
C$_{\mathrm{O}}$ in all the methods. Note that both in GGA+$U_{d}$ and in HSE06, 
the impurity band in the minority spin is clearly separated from the valence band.
The real-space plot of the effective spin density (from GGA) obtained by subtracting 
the charge densities of majority and minority spins 
($\Delta\rho = \rho_{\uparrow}-\rho_{\downarrow}$) is shown in Fig.~\ref{fig:chgplot}. 
One finds that the effective spin density is centered on the C atom and extends to the 
nearest neighbor Zn atoms and the next-nearest neighbor O atoms with a total magnetic moment 
of 2$\mu_{B}$/C. We note that the HSE06 functional gives a  large exchange splitting 
of the impurity states and larger band gap for ZnO than GGA, and the impurity states 
introduced by C$_{\mathrm{O}}$ which mainly consist of carbon states are deep in the gap 
when measured from the valence band maximum (VBM). Therefore, it is not possible to use 
C$_{\mathrm{O}}$ as an acceptor that could introduce hole states for thermally activated 
p-type conductivity in ZnO.  On the other hand, the more stable C$_{\mathrm{Zn}}$ impurity 
acts as a donor and does not lead to any spin polarization, in agreement with Ref.~\cite{Papan-2009}. 

Finally, we discuss possible compensation effects that could alter the magnetic properties 
of C$_{\mathrm{O}}$. Using DFT studies, Li~{\it et~al.} have shown that the presence of 
oxygen vacancy (V$_{\mathrm{O}}$) quenches the spin-polarization from C$_{\mathrm{O}}$ 
in ZnO~\cite{Li-2011}. Comparing the formation energy of neutral defects~\cite{Janotti-2007}, 
we conclude that both, V$_{\mathrm{O}}$ and V$_{\mathrm{Zn}}$, have positive formation energy, 
with the formation energy of V$_{\mathrm{O}}$ being lower than the formation energy of 
V$_{\mathrm{Zn}}$ in O-poor conditions (favorable conditions for C$_{\mathrm{O}}$ as seen 
above). Hence, the role of defects like V$_{\mathrm{Zn}}$ and V$_{\mathrm{O}}$ can be safely 
ruled out in contributing to magnetism in the material. However, the magnetic moment of 
C$_{\mathrm{O}}$ can be quenched by H impurities in ZnO that act as donor. We have tested 
co-doping of H and C in S622 and find that the electron from H compensates one Bohr magneton 
of magnetic moment. Upon adding two H in the supercell, the magnetic moment is completely 
quenched to zero. This observation is independent of the distance of the hydrogen from the 
carbon site in the supercell. In experiments the hydrogen concentration during ZnO synthesis 
is difficult to control. It has been found that the energy barrier for the escape of 
H through the ZnO surface can be as large as 0.58~eV~\cite{Lander-1957}. Thus even if one 
succeeds with C$_{\mathrm{O}}$ doping, one may still fail to obtain magnetic moments due to 
the ubiquitous presence of hydrogen. Similar conclusions have also been derived from 
formation energy studies~\cite{Pham-2011}.

\subsection{\label{IIIB}Fixed spin moment calculations using supercell approach}

\begin{table*}[]
\caption{\label{tab:supercellsinfo}
Nearest neighbor (nn) and next-nearest neighbor (nnn) distance of the periodic image of 
the C impurity in {\AA}, magnetic moments as obtained from VASP (FPLO), and formation energy 
($E_{f}$) for O-poor and C-rich conditions. The total magnetic moment (M$_{\mathrm{Total}}$) 
is shown per supercell, while the magnetic moments of Zn (M$_{\mathrm{Zn}}$) and O 
(M$_{\mathrm{O}}$) are shown per atom, averaged over the nearest neighbor Zn and O, respectively.}
\begin{tabular}{cccccccccccc}
\hline
\hline
\vspace{-3.2mm}\\
 Supercell & r$_{\mathrm{nn}}$~({\AA}) &  r$_{\mathrm{nnn}}$~({\AA}) & M$_{\mathrm{Total}}$~($\mu_{_{\mathrm{B}}}$) & & M$_{\mathrm{C}}$~($\mu_{_{\mathrm{B}}}$) & & M$_{\mathrm{Zn}}$~($\mu_{_{\mathrm{B}}}$) & & M$_{\mathrm{O}}$~($\mu_{_{\mathrm{B}}}$) & & $E_{f}$ (eV) \\
\hline
    S122 & 3.290 & 6.580 & 0.953 (1.194) & & 0.300 (0.743) & & 0.049 (0.072) & & 0.0465 (0.061) & & -4.009 \\
    S221 & 5.284 & 6.580 & 2.000 (2.000) & & 0.591 (1.235) & & 0.118 (0.076) & & 0.0085 (0.081) & & -3.758 \\
    S222 & 6.580 & 10.57 & 2.000 (2.000) & & 0.584 (1.235) & & 0.118 (0.045) & & 0.0068 (0.058) & & -3.910 \\
\hline
\hline
\end{tabular}
\end{table*}

\begin{figure}[]
\centering{
\includegraphics[scale=0.20]{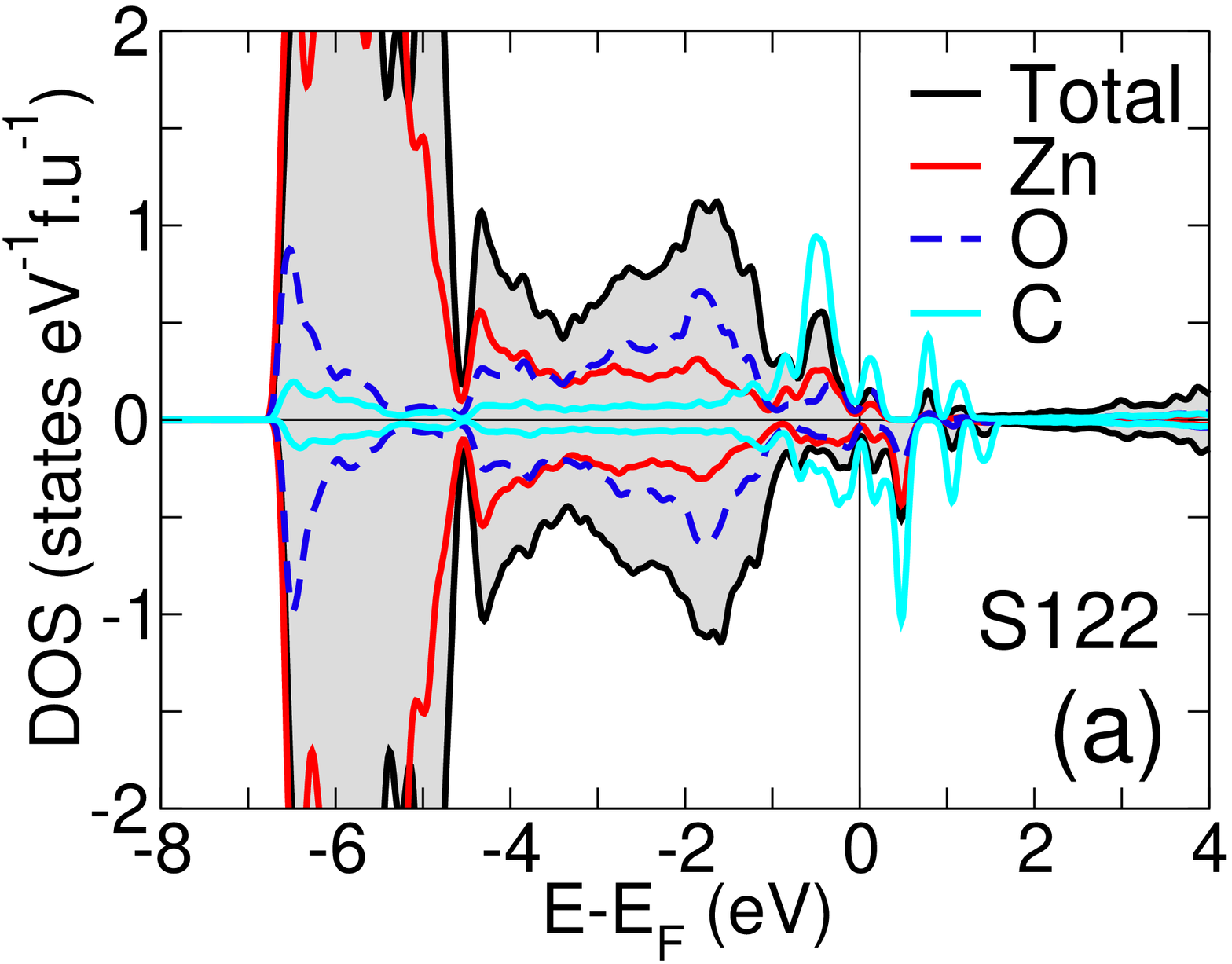}
\includegraphics[scale=0.20]{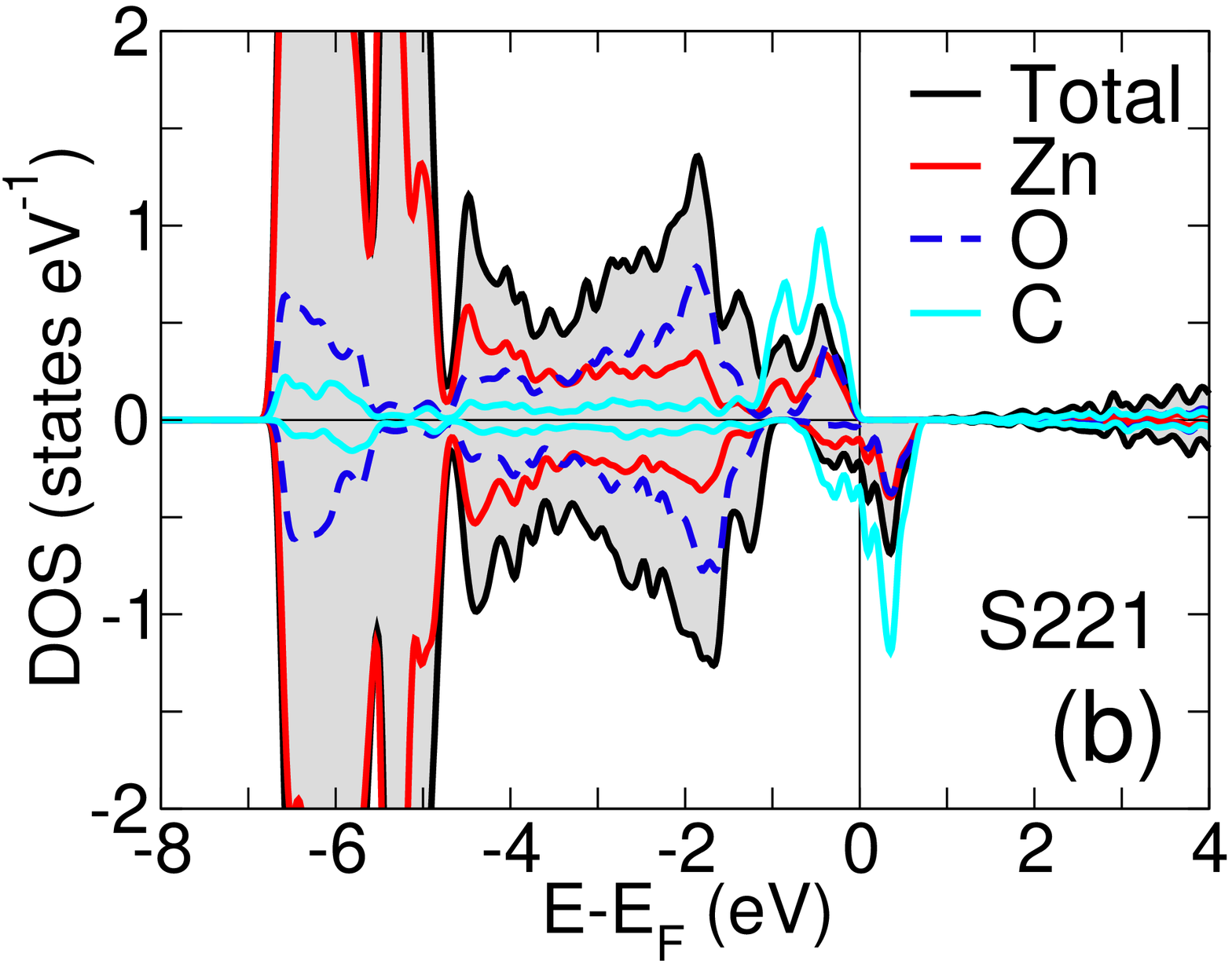}
\includegraphics[scale=0.20]{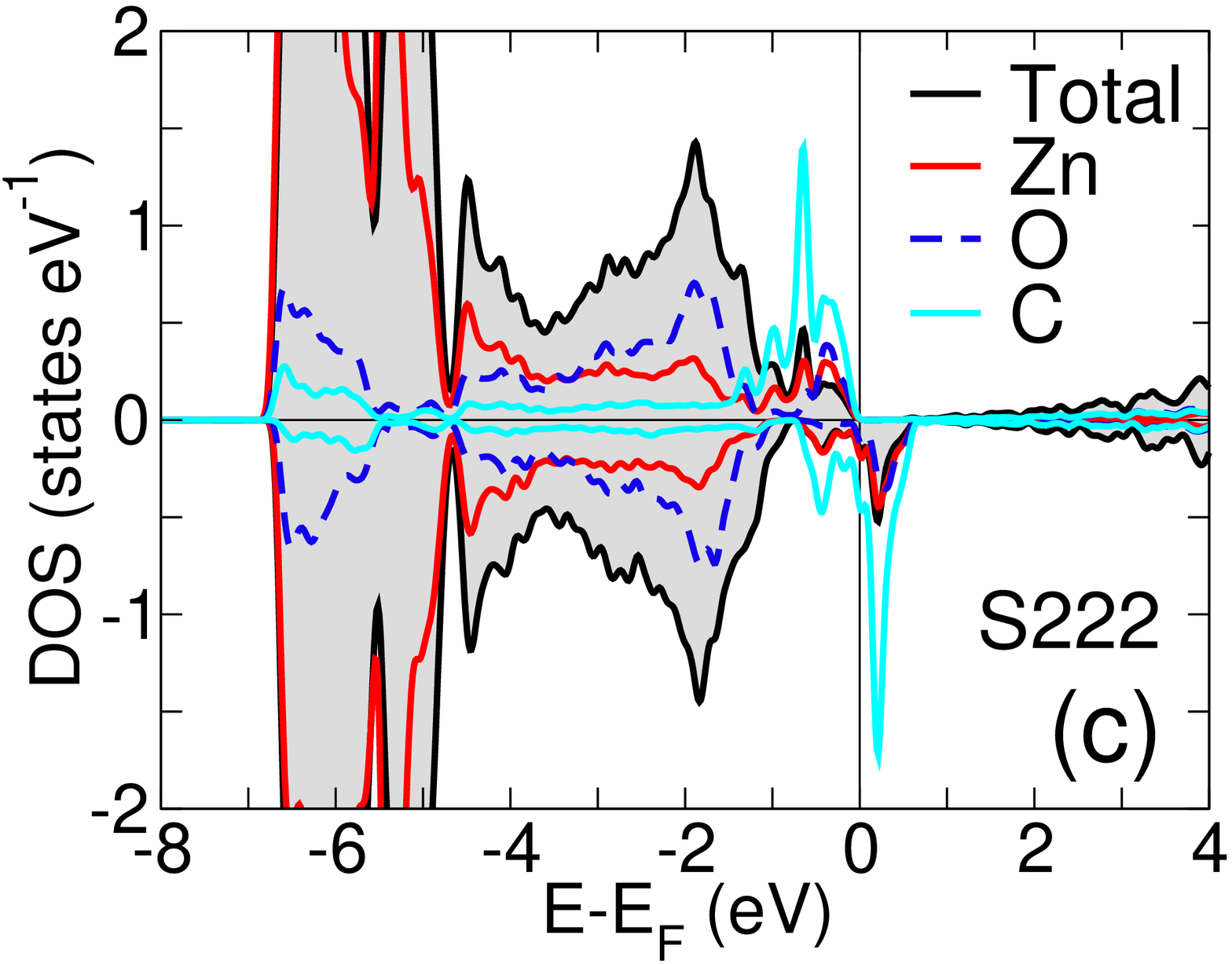}
}
\caption{\label{supercells_dos} (Color online) DOS of C$_{\mathrm{O}}$ in supercells S122 (a), S221 (b) 
and S222 (c), respectively. The DOS consists of the atomic contribution of nearest neighbor Zn and O 
to C (in units of states/atom), while the total DOS is scaled to the number of formula units of ZnO.}
\end{figure}

\begin{figure}[]
\centering{
\includegraphics[scale=0.4]{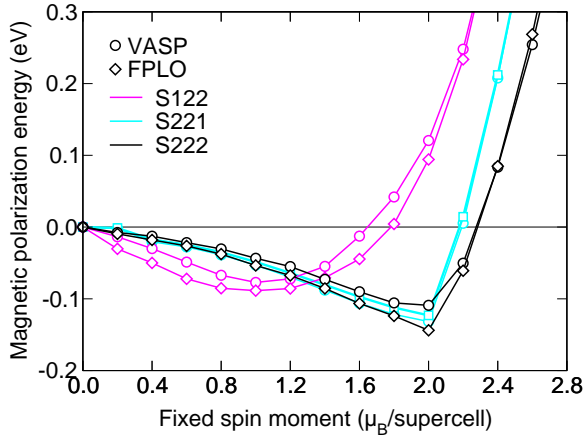}
}
\caption{\label{supercells_fsm1} (Color online) Magnetic polarization energy 
(E$_{\mathrm{sp}}-E_{\mathrm{nsp}}$) (sp = spin~polarized and nsp = non-spin~polarized) as a function 
of the magnetic moment for different supercells from fixed-spin-moment (FSM) calculations. The minima 
in the magnetic polarization energy indicate the most stable magnetic moment in the supercell. 
The magnetic polarization energies obtained from VASP and that of FPLO are in close agreement with each other.} 
\end{figure}

Although the thermodynamic stability considerations of the previous section point to the difficulty 
in incorporating larger amounts of carbon on the oxygen site in ZnO, we continue to discuss the 
consequences of C$_{\mathrm{O}}$ impurities for the magnetic properties, assuming that an appreciable 
concentration can be built up by non-equilibrium preparation techniques. In order to investigate 
impurity interactions at various concentrations, we perform DFT calculations with both VASP and FPLO
using a single C$_{\mathrm{O}}$ impurity in the different supercells shown in 
Fig.~\ref{four_supercells} (a) to (c). These cells correspond to carbon concentrations of 12.5\% 
(for both S122 and S221) and 6.25\% (S222). With decreasing the C concentration from to 12.5\% 
and 6.25\%, the interaction between impurities becomes weaker and the impurity band becomes narrower. 
Even though the supercells S221 and S122 have the same doping concentration of 12.5\%, they do not 
lead to equivalent electronic structure because of the anisotropy of the wurtzite crystal. 
From Table~\ref{tab:supercellsinfo}, we note that the magnetic moment of S122 is a fractional value 
in units of $\mu_{\mathrm{B}}$, an indication of strong hybridization between the C 2$p$ orbitals 
and the valence band, leading to metallic behavior as shown in Fig.~\ref{supercells_dos}(a). 
However, the magnetic moment of S221 and S222 is integer  (2 $\mu_{\mathrm{B}}$), i.e., the system 
is a magnetic half metal (see Fig.~\ref{supercells_dos} (b) and (c)). We carried out 
fixed-spin-moment~(FSM) calculations to test the stability and characteristics of the spin-polarized 
solutions itself. The plot in Fig.~\ref{supercells_fsm1} shows the magnetic polarization energy of 
the supercells shown in Fig.~\ref{four_supercells} (a) to (c). Here the magnetic polarization energy 
is defined as the energy of the system at a given fixed magnetic moment ($m$) relative to the 
energy of non-magnetic solutions ($m = 0$). When a spin-polarized solution is favored, then the magnetic
polarization energy takes negative values. For metallic systems (supercell S122), the minimum magnetic 
polarization energies are found at non-integer values of $m$ as shown in Table~\ref{tab:supercellsinfo}, 
and the magnetic polarization energies show a smooth parabolic change around the minima. For supercells 
S221 and S222 the minimum magnetic polarization energy is obtained at an integer value 
$m$ = 2 $\mu_\mathrm{B}$ and the energy has a cusp at the minima which correspond to localized 
impurity band in the gap. Thus, from the FSM calculations we verify the stability and characteristics 
of spin-polarized solutions in the supercells and one can estimate an upper bound of C concentration 
for the material to be a semiconductor rather than a metal. A higher C concentration leads to a 
metallic solution, i.e.\ the impurity band hybridizes with the valence band of ZnO. According to 
the GGA results, the distance between C impurities should be larger than twice the lattice constant, 
meaning that the C concentration should be lower than 6\%, to avoid such a metallic state. 
The precise estimate of this concentration might depend on the exchange-correlation functional used 
in the calculations. As we discussed above, hybrid-functionals predict the impurity bands to lie 
further up in the gap. Consequently, ZnO may tolerate an even larger C concentration before becoming 
metallic than the values we have estimated from GGA.

\subsection{Exchange interaction in C doped ZnO}
Next, we study the magnetic interaction between the carbon impurities as a function of their separation.
We employ the relatively large supercells S622 and S226~(Fig.~\ref{four_supercells}~d and e) to include 
long-range interactions. The concentration of carbon in these supercells is 4.17\% which is within the 
range of experimental impurity concentrations~\cite{Pan-2007}. As long as the carbon impurities are 
well-separated (maximum separation of 16~{\AA} in $c$-direction and 11~{\AA} in $ab$-direction for 
S226 and S622, respectively), each impurity retains its magnetic moment of $2\mu_B$ independent of 
the distance. Therefore, it is reasonable to map the interaction between two C$_\mathrm{O}$ impurities
onto an effective Ising Hamiltonian
\begin{equation} \label{jijeq}
H(r_{i0}) = -\frac{1}{2}\sum_{\langle i\rangle} J_{i0} (r_{i0}) \sigma_i \sigma_0 ~ ~. 
\end{equation}
The index $i$ runs over the nearest neighbor sites appropriately taking into account the periodic 
boundary conditions used in the DFT calculations. Without loss of generality we can choose 
$\sigma_{0} = \pm 1$ and $\sigma_{i} = 1$ to calculate the individual interaction. For a given geometry 
$r_{i0}=|r_i - r_0|$, the $J_{i0}$ can now be accessed from the total energy difference between the
ferromagnetic and antiferromagnetic alignment of the moments of C$_\mathrm{O}$ impurities. 

\begin{figure}[t]
\centering{
\includegraphics[scale=0.4]{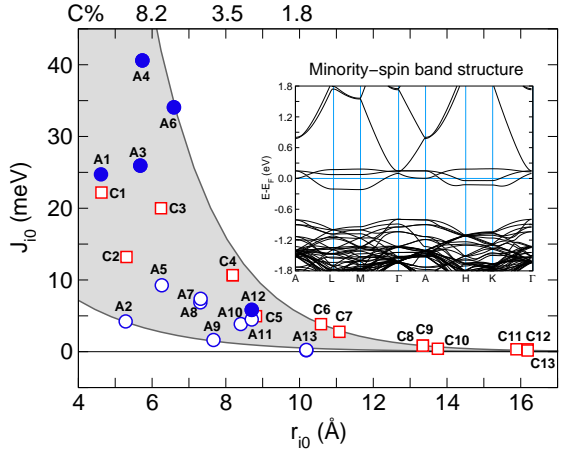}}
\caption{\label{jij-226-622} (Color online) Exchange interaction ($J_{i0}$) extracted from the total 
energy using Eq.~(\ref{jijeq}) plotted against the relaxed C$_{\mathrm{O}}$-C$_{\mathrm{O}}$ distance 
($r_{i0}$). C1~--~C13 and A1~--~A13 represent the different positions of C$_{\mathrm{O}}$ in supercell 
S226 and S622 as in Fig.~\ref{four_supercells}~(d) and (e), with the other C$_{\mathrm{O}}$ positioned 
at C0 and A0, respectively. The range of interaction values is  bounded by two exponential curves. 
The $J_{i0}$ for the C$_{\mathrm{O}}$-C$_{\mathrm{O}}$ oriented along the $ab$-hexagonal plane, which 
are shown as filled symbols, are larger than those along the $c$-axis. From the minority-spin band 
structure of a C$_{\mathrm{O}}$ impurity in ZnO (in GGA) shown in the inset one finds that the 
C$_{\mathrm{O}}$ band crosses the Fermi energy in the hexagonal plane (A--L, $\Gamma$--M, A--H 
and $\Gamma$--K directions) imparting metallic character and thus the magnetic interactions along 
the hexagonal plane are stronger.}
\end{figure}

These $J_{i0}$ are plotted in Fig.~\ref{jij-226-622}. As a guide to the eye, we include a shaded 
area bounded by two exponentially decaying functions within which all the interactions are contained.
We note that the $J_{i0}$ are positive in the whole range, i.e. the spin interaction between 
C$_\mathrm{O}$ is ferromagnetic in the host material. Averagely, the $J_{i0}$ decrease with increasing 
separation $r_{i0}$ between the impurities while there is a large spread in the data for small distances. 
We categorize the geometric configuration of the impurities using the crystallographic direction 
and select those configurations that have the major component of their distance vectors in the 
hexagonal plane ($ab$-direction). The selected configurations (A1, A3, A4, A6 and A12) are shown 
as filled symbols in Fig.~\ref{jij-226-622}. We observe a strongly anisotropic behavior of $J_{i0}$; 
the filled symbols are located in the upper part of the shaded area, whereas the open symbols are 
in the lower part. Test calculations performed for some configurations with the GGA+$U_{d}$ method
have confirmed the ferromagnetic nature of the interactions and indicate that the results are 
independent of the exact energetic position of the minority-spin C 2$p$-states within the host band gap.

Since the carbon impurity levels lie deep inside the band gap (see the DOS in Fig.~\ref{fig:dosplot}~(d)), 
the magnetic interactions cannot be mediated by carriers in the valence or conduction bands of the 
host material (as typical for small-gap DMS materials). Instead, ferromagnetism arises from partly 
filled impurity bands, as in Zener's double-exchange model~\cite{Akai-1998,Dederichs-2005,Mavropoulos-2009}.  
According to this model, ferromagnetic ordering becomes stabilized because it allows for a lowering 
of the kinetic energy of the electrons in the impurity band: Due to the overlap of impurity orbitals of 
the same spin, the dispersion of the impurity bands is increased, and the kinetic energy of electrons 
occupying the lower part of these bands is decreased. Obviously, this energy gain is largest for 
interactions along those crystallographic directions where the impurity band crosses the Fermi level. 

The anisotropic nature of the magnetic interaction can therefore be explained from the band structure 
of C$_{\mathrm{O}}$ in ZnO. In the inset of Fig.~\ref{jij-226-622}, the band structure of the minority 
spin states of C$_\mathrm{O}$ in S333 is shown. Since the spin-majority impurity bands are completely 
filled and thus cannot give rise to any energy gain, only the minority-spin bands 
can be responsible for the magnetic interactions. We find that the carbon bands are crossing the 
Fermi level mostly along the hexagonal $ab$-plane, i.e., along the A--L, $\Gamma$--M, A--H  and 
$\Gamma$--K directions, while along the $c$-axis (M--L, $\Gamma$--A and K--H) the Fermi energy 
lies in the gap between the occupied and the unoccupied bands. Therefore, the ferromagnetic interaction 
is mediated more effectively within the hexagonal $ab$-plane, giving rise to anisotropic ferromagnetism. 

We note that the ferromagnetic interaction is rather short ranged (below 10~\AA). Ferromagnetism at 
finite temperatures can thus only be expected at a carbon concentration larger than 2\% (refer to the 
scale of the C concentration in the upper axis of Fig.~\ref{jij-226-622}). Together with the estimate 
given in Section~\ref{IIIB}, which gives the upper bound for C concentration to avoid metallic state, 
we conclude that a C$_\mathrm{O}$ concentration between 2\% and 6\% should be optimal to obtain 
ferromagnetism due to homogeneously distributed impurities. Moreover, we note that C impurities have 
a tendency to cluster which poses an upper limit to the useful concentration. As pointed out in previous 
work~\cite{Wu-2010}, the C impurities tend to form energetically more stable C$_2$ molecules. 
In this case, the $pp\pi^*$ orbital of the C$_2$ molecule resonates with the conduction band of ZnO 
which may lead to ferromagnetism mediated by the host material in n-type ZnO. Thus, this mechanism 
differs from the ferromagnetic interaction between C$_\mathrm{O}$ impurities described in the present 
work which originates from conductivity due to impurity bands, and is hence independent of host carriers. 
 
\section{Conclusions}

We have studied the two possible substitutional carbon impurities in ZnO (C$_{\mathrm{O}}$ and 
C$_{\mathrm{Zn}}$) and find that C$_{\mathrm{Zn}}$ is energetically more favorable than 
C$_{\mathrm{O}}$. However, C$_{\mathrm{O}}$ can be stabilized under specific (O-poor and C-rich) 
growth conditions. This type of environment is uncommon in usual ZnO growth, but could be generated 
by oxidizing metallic zinc in the presence of an atomic carbon source. Given that a material with 
a C$_{\mathrm{O}}$ concentration of 2--6\% can be realized, this would be an interesting prototypical 
system for $d^{0}$ magnetism: Our DFT calculations indicate that C$_{\mathrm{O}}$ in this range 
of concentrations is associated with a localized magnetic moment of 2~$\mu_{\mathrm{B}}/\mathrm{C}$. 
Partially filled impurity bands in the minority spin channel mediate ferromagnetic interaction 
between the C$_{\mathrm{O}}$ impurities. These interactions are short-ranged and anisotropic, 
being stronger within the $ab$-plane of the wurtzite ZnO crystal than along the $c$-axis. 
Based on our calculations, we predict that such layered ferromagnetism could be used as an 
experimental hallmark to distinguish carbon-induced magnetism from possible other forms of 
defect-induced ferromagnetism in ZnO.

\section*{Acknowledgments}

We thank Andreas Ney and Heike Herper at University of Duisburg-Essen for fruitful discussions. 
We acknowledge the computational resources from Center for Computational Sciences and Simulation (CCSS), 
University of Duisburg-Essen. The work was funded by Deutsche Forschungsgemeinschaft through 
the Research Trainig Group GRK-1240 ``Photovoltaics and Optoelectronics from Nanoparticles". 


\end{document}